\documentclass[12pt]{article}
\usepackage[dvips]{epsfig}
\usepackage{amssymb}

\newcommand{\beq}{\begin{equation}}
\newcommand{\eeq}{\end{equation}}
\newcommand{\bq}{\begin{quotation}}
\newcommand{\eq}{\end{quotation}}
\newcommand{\bc}{\begin{center}}
\newcommand{\ec}{\end{center}}
\newcommand{\BFACE}[1] {\mbox{\boldmath $#1$} }

\begin{document}

\title{{\vspace{0cm}
\addtocounter{footnote}{1}
\sc Weyl's principle, cosmic time and quantum 
fundamentalism\footnote{To appear in the section on ``Physical and philosophical perspectives on probability and time" in S. Hartmann {\em {et al.}} (eds.) {\em{Explanation, Prediction and Confirmation}}, Springer's The Philosophy of Science in a European Perspective book series.}}}

\author{
\addtocounter{footnote}{-2}
{\sc S.E. Rugh\footnote{Symposion, 
`The Socrates Spirit', Section for Philosophy and the Foundations of Physics,
Helleb\ae kgade 27, Copenhagen N, Denmark
({\em e-mail: rugh@symposion.dk)}} 
\addtocounter{footnote}{5}
and  H.
Zinkernagel\footnote{Department of Philosophy I, Granada University, 18071
Granada, Spain ({\em e-mail: zink@ugr.es}).}} }
\date{}

\maketitle

\begin{abstract}
\noindent
We examine the necessary physical underpinnings for setting up the cosmological 
standard model with a global cosmic time parameter. In particular,
we discuss the role of Weyl's principle which asserts that cosmic matter 
moves according to certain regularity requirements.
After a brief historical introduction to Weyl's principle
we argue that although the principle is often not explicitly mentioned 
in modern standard texts on cosmology, it is implicitly assumed
and is, in fact, necessary for a physically well-defined notion of cosmic time. 
We finally point out that Weyl's principle might be in conflict with 
the wide-spread idea that the universe at some very early stage can be 
described exclusively in terms of quantum theory.
\end{abstract}

\section{Introduction}

A basic characteristic of the 
Friedmann-Lema\^{\i}tre-Robertson-Walker (FLRW) model 
is its $t$ parameter which is employed by cosmologists to trace back
the evolution of the universe to its early stages.
In a previous examination, we defended
a `time-clock' relation which asserts that time, 
in order to have a physical basis, must be understood in relation to physical 
processes which act as `cores' of clocks (Rugh and Zinkernagel 2009). 
In particular, we argued that a necessary physical condition for {\em interpreting} 
the $t$ parameter of the FLRW model as cosmic time in some `epoch' of the universe 
is the (at least possible) existence of a physical process which 
can function as a core of a clock in the `epoch' in question.\footnote{One of 
our results was that there are interesting problems for making 
this $t \leftrightarrow \mbox{time}$ interpretation, and thus establishing 
a physical basis for cosmic time (in particular for a cosmic time {\em scale}), 
at least at $\sim 10^{-11}$ seconds after the ``big bang" -- 
that is, approximately 30 orders of magnitude before 
(in a backwards extrapolation of the FLRW model) Planck scales are reached.} 

In this paper we shall argue, in conformity with -- but independently of -- 
the time-clock relation, that the very set-up of the standard (FLRW) model in cosmology with a 
global time is closely linked to the motion (and properties) of cosmic 
 matter.\footnote{Whereas our aforementioned study
examines the physical basis for time both locally and globally we shall assume
in the present manuscript that spacetime is physically well-defined locally.} 
It is often assumed that the FLRW model 
may be derived just from the {\em cosmological principle} which states 
that the universe is spatially homogeneous and isotropic (on large scales). 
It is much less well 
known that another assumption, often called Weyl's principle, is necessary -- or, at 
least, have been claimed to be necessary -- in order to arrive at the FLRW 
model and, in particular, its cosmic time parameter.  In a version close to 
Robertson's (1933) (we shall discuss various formulations later), the principle states:
\begin{quotation}
\noindent
{\em Weyl's principle}: The world lines of galaxies, or `fundamental
particles', form (on average) a spacetime-filling family of non-intersecting geodesics 
converging towards the past.
\end{quotation}

The importance of Weyl's principle is that it provides a reference 
frame based on an expanding `substratum' of `fundamental particles'.
In particular, if the geodesic world lines are required to be orthogonal to a series 
of space-like hypersurfaces, a comoving reference frame is defined in 
which constant spatial coordinates are ``carried by" the fundamental particles. 
The time coordinate is a cosmic time which labels the series of hypersurfaces, and 
which may be taken as the proper time along any of the particle world lines.

Insofar as the Weyl principle is necessary for the notion of cosmic time in the FLRW model, 
it clearly becomes important to examine whether the properties and 
motion of matter are compatible with the Weyl principle as we go back in 
cosmic time.\footnote{A related question may of course be made concerning 
the cosmological principle, see e.g. Weinberg (1973, p. 407).}  
If a point is reached at which this is not the case,  then it appears 
not to be physically justified to contemplate `earlier' epochs. Doing so 
would involve extrapolating the FLRW model into a domain where the fundamental 
assumptions (needed to build up the model) are no longer valid and the model 
would lose its physical basis (see also the discussion in Rugh and 
Zinkernagel 2009, p. 5).

In the following, we first briefly review the early history of Weyl's principle and 
question a claim, found in some of the recent literature on this principle, 
to the effect that the principle has been replaced by the cosmological
principle. We then show that although the Weyl principle
 is not often mentioned explicitly in modern texts on cosmology, it is 
nevertheless in these texts in implicit form (and, we argue, necessarily so). 
We finally discuss and question the prospect of satisfying Weyl's principle, and hence define 
cosmic time, at a very
`early phase' of the universe, if this phase is thought to be describable
exclusively in terms of quantum theory.

\section{A very brief history of Weyl's principle}

The early history and reception of Weyl's principle -- sometimes denoted postulate, 
assumption or hypothesis --  have been chronicled e.g. by North 1990, and more recently by 
Bergia and Mazzoni (1999) and Goenner (2001), see also 
Ehlers (2009). In this section we sketch a few important points about the 
historical development, and we argue against an apparent consensus among 
Bergia and Mazzoni (1999) and Goenner (2001) which takes Weyl's principle to have been 
rendered redundant by the cosmological principle.

Weyl first introduced his principle in 1923, with the appearance of the 
5th and revised version of his {\em Raum, Zeit, Materie}, in connection with a 
discussion of de Sitter's solution to Einstein's field equations.
In Weyl's 1926
formulation of his principle (which he here called hypothesis), it reads:\footnote{Weyl's 
formulations in his 1923 writings are more convoluted, see Goenner (2001) and
 Bergia and Mazzoni (1999).}
\begin{quote}
\dots the world lines of the stars [in later contexts, galaxies] form a sheaf [bundle], 
which rises in a given direction from the infinitely distant past, 
and spreads out over the hyperboloid [representing de Sitter's model] in the 
direction of the future, getting broader and broader. [Quoted from Goenner (2001, p. 121), 
our inserts]
\end{quote}
This principle, or hypothesis, amounts to specifying a choice of congruence (a 
family of non-crossing curves which fills spacetime) of timelike geodesics to 
represent the world lines of the cosmological substratum, see e.g. Goenner (2001, p. 120). 
As Weyl emphasized, such an assumption concerning the choice of congruence is necessary 
to derive an unambiguous cosmological redshift in de Sitter's model; see also Bergia 
and Mazzoni (1999, pp. 336-338). As for the possible empirical support of 
his principle, Weyl mentions (6 years before Hubble's 1929 paper) that ``it appears that the velocities between distant celestial objects on average 
increase with their mutual separations" (quoted from Ehlers 2009, p. 1655).

Weyl (1923, p. 1664) notes that on his hypothesis the stars (galaxies)
belong ``to the same causally connected system with a common
origin".\footnote{According to Ehlers (2009, p. 1657) this ``implies that
each particle can be influenced by all others at any time; in modern
parlance there is no particle horizon".} Moreover, Weyl hints (same page)
that this causal connectedness of the stars implies an assumption of the
`state of rest' of the stars which is ``the only one compatible with the
homogeneity of space and time". On this point, Goenner
(2001, p. 120) comments that Weyl thus ``indicates that his
hypothesis implies the existence of a common time parameter or, turned
around, that the stars have a common instantaneous rest system". While
the implication of a cosmic time is consistent with Robertson's use of
Weyl's principle (see below), Goenner also notes (p. 126) that Weyl never explains in detail
why his chosen congruence also implies a common rest frame for the
galaxies (and thus a cosmic time).

As we shall see in section 3, the issue of whether or not a common time 
(a common rest frame) is implied by Weyl's principle may be responsible for 
the difference in formulations of this principle 
in the literature. While Robertson 
is somewhat ambiguous concerning what is included in 
the Weyl principle (see section 3), he notes (1933, p. 65) that the 
reintroduction in cosmology of a significant simultaneity (a cosmic time) 
implied by Weyl's postulate is permissible since observations support the idea
 that galaxies (on average) are
moving away from each other with a mean motion which represents the
actual motion to within relatively small and unsystematic 
deviations.\footnote{Robertson's empirical justification for the 
introduction of a cosmic time stands in contrast to Friedmann's statement 
(1922, p. 1993): ``In the expression for $ds^2$, $g_{14}$, $g_{24}$, $g_{34}$ can be 
made to vanish by corresponding choice of the time coordinate, or, shortly said, 
time is orthogonal to space. It seems to me that no physical or philosophical reasons 
can be given for this second assumption; it serves
exclusively to simplify the calculations".}

Whereas, for Weyl, the selection of a particular congruence of curves as 
world lines to represent cosmic matter was originally merely a specific 
property of the de Sitter universe, it later became (e.g for  
Robertson) a necessary assumption for constructing cosmological models. This role of 
Weyl's principle is emphasized by Ehlers in Bertotti (1990, p. 29); see also 
Weyl (1930, p. 937):
\begin{quote}
H. Weyl in 1923 points out that to have a 
cosmological model one {\em has
to specify}, besides a space-time (M,g), a congruence of timelike curves to
represent the mean motion of matter. [Our emphasis]
\end{quote}

Now, Weyl introduced his principle in de Sitter space where 
(unlike the FLRW model) there is no unique choice of congruence 
(this is why de Sitter's cosmos can be written either as a static or an expanding universe), 
see e.g. Ellis (1990, p. 100). But, as we shall discuss further in section 3 and 4, even 
if the choice of congruence is unique in the FLRW model, it is still 
crucial that the actual matter content of the universe is (on average) well represented by this congruence.

Bergia and Mazzoni (1999, p. 339) note: ``In his 1929
paper, Robertson had given no justification for his introduction of a
cosmic time. As we have just seen, he did offer some in 1933, guided by
Weyl's principle. Therefore the continuity between Weyl's and the
cosmological principle seems fairly well established". This quote 
might indicate that the cosmological principle somehow replaced the
Weyl principle but such an idea 
would, in our assessment, be misleading both for historical and conceptual 
reasons.\footnote{The term ``cosmological principle" 
is due to Milne in 1933, though it can already implicitly be found in 
Einstein's 1917 paper, see e.g. North (1990, p. 157).}
For Robertson (1933, p. 65) clearly states that he uses two (in fact four) 
assumptions amounting to {\em both} Weyl's principle and the 
cosmological principle.\footnote{The four assumptions, schematically, 
are: (1) a congruence of geodesics; (2) hypersurface orthogonality; (3) homogeneity; and (4) isotropy.
In section 3 we return to the relation between (1), (2) and 
Weyl's principle in Robertson (1933).} 
Furthermore, Robertson notes that the cosmic time 
implied by Weyl's principle ``allows us to give a relatively 
precise formulation of the assumption that our ideal 
approximation to the actual world is spatially uniform" (1933, p. 65), and he
 thus suggests that the Weyl principle
 is actually a {\em precondition} for the cosmological principle (we shall pursue
 this theme further in section 3).

In a spirit which seems similar to that of Bergia and Mazzoni,  
Goenner (2001, p. 126) notes in connection with the literature of the 1940s: 
``In fact, Weyl's hypothesis had become superfluous and was 
replaced by the {\em cosmological principle}, i.e. the hypothesis 
that, in the space sections, no point and no direction are preferred". 
Apparently, Goenner's assessment of the (ir-)relevance of Weyl's principle for today's 
cosmology is similar: ``Weyl's stature in mathematics and science may \dots
explain why the hypothesis still is mentioned in some modern 
books on gravitation and cosmology, notably by authors not 
specialized in cosmological research" (2001, p. 127). This statement fits well 
with the fact that Weyl's principle is often absent in an explicit form 
in current cosmology textbooks. However, as we shall argue in the following 
section, Weyl's principle is, at least implicitly, still present (and necessarily so) 
 in the main texts on cosmology.

\section{Weyl's principle in standard texts on cosmology}

In some cosmology textbooks, e.g. Bondi (1960), Raychaudhuri 
(1979) and Narlikar (2002), the importance of Weyl's principle is emphasized, and explicitly 
referred to, when the physical basis of the comoving frame, cosmic time, 
and the FLRW model are outlined. For instance, the derivation of the FLRW metric 
in Narlikar (2002, p. 107 ff.) is explicitly built on two assumptions, namely:
\begin{enumerate}

\item Weyl's postulate (Narlikar): The world lines of galaxies [or `fundamental
particles'] form a bundle of non-intersecting geodesics orthogonal to a
series of spacelike hypersurfaces.

\item The cosmological principle: The universe, on large scales, 
is spatially homogeneous and spatially isotropic.\footnote{We note that while Weyl's principle and the cosmological principle allow for the possibility 
to set up the FLRW model with a {\em global} cosmic time, the 
implementation of these principles can only be
motivated physically if we already have a physical foundation for the
concepts of `space' and `time' {\em locally} -- otherwise we cannot
apply concepts like ``spacelike", ``spatially homogeneous", ``spatially
isotropic", which appear in the definitions of these principles.}
\end{enumerate}

\noindent 

In Narlikar's formulation of Weyl's postulate (which includes the 
orthogonality criterion; see below), this 
postulate is sufficient to build up a comoving reference frame in which the 
constituents of the universe are at rest (on average) relative to the comoving 
coordinates: The trajectories, $x_i = \mbox{constant}$, of the constituents 
are freely falling geodesics, and the requirement that the geodesics be orthogonal 
to the spacelike hypersurfaces
translates into the requirement $g_{0i} = 0$, which (globally) resolves 
the space-time into space and time (a 3+1 split). We have $g_{00} = 1$ if we 
choose the time coordinate $t$ so that it corresponds to proper time 
($dt = ds/c$) along the lines of constant $x_i$, i.e. $t$ corresponds to 
clock time for a standard clock at rest in the comoving coordinate 
system.\footnote{That the world lines are geodesics implies that $g_{00}$ depends 
only on $x_0$,  and so that $g_{00}$ can be set to unity by a 
suitable coordinate transformation, see e.g. Narlikar (2002, p. 109).} 
The metric can thus be written in {\em synchronous} form in which the spacelike
 hypersurfaces are surfaces of simultaneity for the comoving observers 
(see also e.g. MTW 1973, p. 717):
\begin{equation} \label{synchronous}
ds^2 = c^2dt^2 - g_{ij}dx^idx^j  \; \; \; \; i, j = 1,2,3.
\end{equation}

In Narlikar (2002), the role of the cosmological principle 
is then to simplify the spatial part of this metric in order to get 
the standard FLRW form:\footnote{Although there is a preferred choice of 
the congruence of world lines (a preferred reference frame)
in the FLRW model (see below), 
there are many different coordinate representations of this model;
see e.g. Krasinski (1997, p. 11 and p. 14 - 16) who outline at least 
five different coordinate representations. }
\begin{equation} \label{FRLWlineelement}
ds^2=c^2dt^2 - R^2(t)\left\{ \frac{dr^2}{1-kr^2} + r^2 (d\theta ^2 +
\sin ^2 \theta d\phi ^2) \right\}
\end{equation}

Narlikar's discussion of the assumptions going into the derivation of 
the FLRW line element seems to follow Robertson (1933) closely with 
one notable difference: Narlikar takes orthogonality of 
the matter world lines to the series of space-like hypersurfaces as 
{\em being part of} Weyl's principle (and hence, implicitly, that the 
existence of a comoving frame follows directly from this principle). By contrast, 
Robertson states Weyl's principle as in section 1 (i.e. as the principle
that matter is 
represented by a congruence of diverging geodesics) and add, as a {\em further} assumption, 
that the space-like hypersurfaces are orthogonal to the 
congruence of geodesics.\footnote{However, Robertson is actually somewhat ambiguous about 
what exactly is included in, and implied by, Weyl's principle. For, right 
after introducing 
{\em both} assumptions (congruence and hypersurface orthogonality), he mentions (1933, p. 65): ``The possibility 
of thus introducing in a natural and significant way this {\em cosmic time t} we consider 
as guaranteed by Weyl's postulate, which is in turn a permissible extrapolation from 
the astronomical observations". Perhaps this 
ambiguity is related to Weyl's own insufficient explanation, mentioned 
in section 2, of whether a comoving frame (`state of rest' of the stars) follows
directly from his principle.} 
In any case, whether or not the assumption of hypersurface orthogonality
is included in Weyl's principle, it is clear that one can impose the
requirement of the congruence being orthogonal to the hypersurfaces only
given that there {\em is} a congruence.\footnote{One can have a congruence
which is not orthogonal to a series of spacelike hypersurfaces but not,
of course, a hypersurface orthogonal congruence which is not a
congruence! Note however the underlying coupled problem: The specification of a congruence (is it hypersurface orthogonal?, are the world lines geodesics?, etc.) depends on $g_{\mu\nu}$, and the specific form of $g_{\mu\nu}$ depends in turn on the choice of congruence (the reference frame).}

Given the importance of Weyl's principle in Robertson (1933), it may at 
first be surprising that other (than the above mentioned) standard text books 
on general relativity and modern cosmology, such as Misner, Thorne and Wheeler 
(MTW) (1973), Wald (1984), Peebles (1993) or Hawking and Ellis (1973), have 
no explicit reference to Weyl's principle. As far as we can see, however, the reason is 
simply that the Weyl principle is assumed implicitly in these books at an 
early stage of setting up the FLRW model. To see this, consider first Ellis' 
(1990, p. 99) clarification (cf. also the quote of Ehlers 
in section 2 above):
\begin{quote}
It is important to realise that a cosmological model is specified only 
when a 4-velocity $u^a$ representing the average motion of matter in the 
universe has been specified as well as the space-time metric $g_{{\mu}{\nu}}$; 
the observable relations in the model are determined by the choice of this 
4-velocity, or equivalently of the associated fundamental world-lines.
\end{quote}

As mentioned in section 2, the preferred choice of a 4-velocity or, equivalently, 
a congruence of world lines, to represent the average motion of matter is 
unique in the FLRW case.\footnote{The mentioned equivalence follows if 
precisely one non-zero 4-velocity vector is assigned to every point of the manifold (the world line is obtained as the integral curve of $u^a$).  Ehlers (1961) 
introduces a three-dimensional family of timelike curves to represent the
world lines of the matter elements with respect to an 
arbitrary local coordinate system, $x^{a} = x^{a}( y^{\alpha}, s) $.
He then derives the 4-velocity of the matter elements
by differentiating the coordinates $x^a$ (i.e.
$ u^a = \partial x^a/ \partial s  $) with respect to the proper
time $s$ along the world lines. Subsequently 
the acceleration $\dot{u}$ of the matter elements
can be defined, and this acceleration vanishes ($\dot{u} = 0$) if 
the world lines move along geodesics.}
But still, the congruence plays a fundamental role 
since the symmetry constraints of homogeneity and isotropy are 
imposed {\em with respect to} such a congruence, cf. e.g. Ellis (1999):
\begin{quote}
We start by assuming large-scale spatial homogeneity 
and isotropy {\em about a particular family of worldlines}. The RW models 
used to describe the large-scale structure of the universe
embody those symmetries exactly in their geometry. It follows that 
comoving coordinates can be chosen \dots . [Our emphasis] 
\end{quote}
Another way of stating this point is that isotropy can be satisfied 
only in a particular reference frame or for a particular class of 
fundamental observers (other observers moving with respect to these will not see isotropy). 
Indeed, such a fundamental class of observers (congruence) is 
part of the definition of isotropy, cf. e.g. Wald (1984, p. 93) (see also MTW 1973, p. 714):
\begin{quote}
\noindent 
A spacetime is said to be (spatially) {\em isotropic} at 
each point if there exists a congruence of timelike curves (i.e. observers), 
with tangents denoted $u^a$, filling the spacetime \dots [such that] \dots it is 
impossible to construct a geometrically preferred tangent vector orthogonal to $u^a$.
\end{quote}
Thus, Weyl's principle -- in the general sense of matter being well represented by a congruence of world lines -- is a precondition for the cosmological principle; the former can be satisfied without the latter being satisfied but not vice versa.

As hinted above in connection with Robertson, 
the specification of a congruence of world lines representing matter is a necessary but 
not sufficient condition for setting up a 
cosmic time.\footnote{For instance, G\" odel's 
expanding and rotating solution from 1952 satisfies, as far 
as we can see, Weyl's principle in Weyl's 1926 (and Robertson's 1933) formulation. 
But this solution does not have a cosmic time (at least not for high rates 
of rotation, see  e.g. Belot (2005, p. 27)). See also Ellis (1996) for a detailed discussion
 both of G\" odel's static and expanding rotating solutions of Einstein's field equations.}
Only with the additional requirement of the congruence of world lines 
being hypersurface orthogonal do we get a 
sufficient condition.
In terms of the 4-velocity field, a sufficient condition for having cosmic time is 
that the motion of matter is (on average) described by this field and that the motion 
is irrotational (corresponding to the 4-velocity having 
zero vorticity; see e.g. Ellis (1996).\footnote{The 4-velocity field may be decomposed 
into rotation (vorticity), shear and expansion components, see e.g. 
Ehlers (1961, p. 1228) or MTW (1973, \S 22.3). As concerns the connection 
between zero vorticity and hypersurface orthogonality, Malament (2006, p. 251) 
presents a nice picture: ``Think about an ordinary rope. In its natural twisted state, the rope 
cannot be sliced by an infinite family of slices in such a way that each slice is orthogonal 
to all fibers. But if the rope is first untwisted, such a slicing is possible. Thus 
orthogonal sliceability is equivalent to fiber untwistedness. The proposition 
extends this intuitive equivalence to the four-dimensional `spacetime ropes' 
(i.e. congruences of worldlines) encountered in relativity theory. It asserts that 
a congruence is irrotational (i.e. exhibits no twistedness) iff it is, at least locally, 
hypersurface orthogonal."} 
Asserting the existence of the 
4-velocity field (representing matter) is of course prior
to inquiring about its vorticity. This is just another way of repeating that 
the condition that matter can be well described by Weyl's principle is necessary 
for having cosmic time.

To require that the motion of matter is well represented by a 
congruence of world lines (i.e. to impose Weyl's principle in the general sense) 
is to require that 
the matter world lines are {\em non-crossing} (of course, this can only be true 
on average, see below). This non-crossing of world lines is built into the 
construction of the comoving frame with respect to which cosmic time is defined. 
As described e.g. in MTW (1973, p. 715 ff), see also Wald (1984, p. 95) and 
Weinberg (1972, p. 338), an arbitrary grid of space coordinates $(x^1, x^2, x^3)$ 
(constant labels) are laid out on a spacelike hypersurface (of homogeneity). 
These coordinates are ``propagated off" and throughout all spacetime {\em by means of} the 
world lines of the cosmological fluid with proper (= cosmic) time measured along any of the 
fluid world lines (so the coordinates are ``carried by" the world lines). 
Since the world lines of the cosmological fluid are used to propagate the coordinates
 it is crucial that there is {\em no crossing}
of the world lines (i.e. that the family of world lines constitutes 
a congruence), as we would otherwise have the same spacetime point described 
by different (incompatible) coordinates.\footnote{Note that in Minkowski 
spacetime there is no unique congruence of world lines (no unique preferred frame), 
no (preferred) cosmic time, and no need to impose the non-crossing criterion -- 
but also that one does not need the world lines of the material constituents 
to propagate (set up) the coordinates. It is however possible to do so, 
and in that case it {\em is} necessary that the world lines are non-crossing, 
see e.g. Peebles (1993, p. 250).} 
As Narlikar puts it (2002, p. 108):
\begin{quote}
It is worth emphasizing the importance of the non-intersecting 
nature of world lines. If two galaxy world lines did intersect, our [comoving] 
coordinate system above would break down, for we would then have two different 
values of $x^{\mu}$ specifying the same point in spacetime 
(the point of intersection).
\end{quote}

How can Weyl's principle be fulfilled in the {\em real} universe?  Typical ordinary 
velocities of (nearby) galaxies relative to each other
are $<v> \; \sim 1/1000 \times c$ (MTW 1973, p. 711) and, indeed, some
galaxies do collide. Likewise with the more fundamental constituents in earlier phases of the universe.
Thus the fundamental world lines in the Weyl principle must be 
some `average world lines' associated with the average motion
of the fundamental particles (in order to ``smooth out" any crossings).\footnote{A closely related 
problem is to average out inhomogeneities in the matter distribution (such averaging procedures have
been developed to a large degree of sophistication, see e.g. Krasinski (1997, pp. 263 - 275)). 
It is a highly non-trivial problem, and it was emphasized already 
by G\" odel (1949, p. 560), that the necessary averaging over large volumes will introduce an 
arbitrariness in the definition of cosmic time depending on
the details of the averaging process and the size of the regions considered 
(see also North (1990, p. 360) and Dieks (2005, p. 11 - 12)).}

At present and for most of cosmic history, the comoving frame of reference can 
be identified as the frame in which the cosmic microwave background radiation 
looks isotropic (see e.g. Peebles 1993, p. 152), and cosmic matter is 
(above the homogeneity scale) assumed to be
described as dust particles with zero pressure which fulfill Weyl's principle. 
In the early radiation phase, matter is highly relativistic (moving with 
random velocities close to $c$), and the Weyl principle is not satisfied for a typical 
particle but one may still introduce fictitious averaging volumes in order to 
create substitutes for `galaxies which are at rest'; 
see e.g. Narlikar (2002, p. 131). 

However, above the electroweak phase transition (before $10^{-11}$ seconds 
`after' the big bang), all constituents are massless and move with 
velocity $c$ in any reference frame. There will thus be no constituents which are comoving  
(at rest).\footnote{This conclusion may also be reached by noting that the set-up of 
the FLRW model requires matter (the energy-momentum tensor) to be in the form of 
a perfect fluid, as this is the only form compatible with the FLRW symmetries, 
see e.g. Weinberg (1972, p. 414). And a source consisting of pure radiation is not 
sufficient since one cannot effectively simulate a perfect fluid by ``averaging over 
pure radiation": Krasinski (1997, p. 5 - 9) notes
that the energy-momentum tensor in cosmological models may contain many different contributions, 
e.g. a perfect fluid, a null-fluid, a scalar field, and an 
electromagnetic field. But he also emphasizes that whereas a scalar field source is 
compatible with the FLRW geometry (since it acts as  a stiff perfect fluid with 
equation of state $p = \rho$), a source of pure null fluid or pure 
electromagnetic field is not compatible with the FLRW geometry, and solutions with 
such energy-momentum sources have no FLRW limit (see Krasinski 1997, p. 13).} 
One might attempt to construct mathematical points (comoving with a reference frame) 
like the above mentioned center of mass (or, in special relativity, center of energy) 
out of the massless, ultrarelativistic gas
particles, but this procedure requires that length scales be available in order to 
e.g. specify how far the particles are apart (which is needed as input in the 
mathematical expression for the center of energy). As discussed in Rugh and 
Zinkernagel (2009) the only option for specifying such length scales (above the
electroweak phase transition) will be to 
appeal to speculative physics, and the prospects of satisfying Weyl's principle
(and have a cosmic time) will therefore 
also rely on speculations beyond current well-established physics.

We conclude that it is instrumental that some averaging procedure is made in order 
to yield a non-crossing family of world lines (a congruence). Whether this 
is possible when matter is described by quantum theory (e.g. in the
 very early universe) is the question we address in the next section.

\section{Cosmic time with quantum matter?}

We have seen that Weyl's principle cannot be disregarded in the FLRW model 
as it is either implicitly or explicitly included among the fundamental principles
used to set up this model. The question therefore arises: What could
be candidates for the ``Weyl substratum" which, at epochs when no galaxies are present,  
can form substitutes for (on average) non-intersecting galaxies at rest? 

The empirical adequacy
of both Weyl's principle and the cosmological 
principle depends on the actual arrangement and motion of
the physical constituents of the universe. As we go backwards in time it may become 
increasingly difficult to satisfy these
physical principles since, as mentioned in section 3, the nature of the physical
constituents is changing from galaxies, to relativistic gas particles,
and to entirely massless particles moving with velocity
$c$. In particular, the Weyl principle refers to a 
non-crossing family of (fluid or particle) world lines, that
is, to classical or classicalized particle-like behavior of the material constituents.
This makes it difficult even to formulate the Weyl  principle (let alone decide whether 
it is satisfied) if some period in cosmic history is reached where the
`fundamental particles' are to be described by wave-functions $\psi (x,t)$
referring to (entangled) quantum constituents. What is a `world line' or
a `particle trajectory' then? Unless one can specify a 
clear meaning of 'non-intersecting trajectories' in a contemplated quantum 
`epoch' it would seem that the very notion of cosmic time, and hence the 
notion of `very early universe' is compromised.

This last problem of identifying a Weyl substratum within a 
quantum description arises most clearly on a ``quantum fundamentalist" view 
according to which the material constituents of the universe could be 
described {\em exclusively} in terms of quantum theory at some early 
stage of the universe.\footnote{For instance, Kiefer notes that ``The Universe was 
essentially `quantum' at the onset of inflation" (Joos et al. 2003, p. 208).}
On such a quantum fundamentalist view, the following question naturally arises
\begin{quotation}
\noindent
{\em The cosmic measurement problem}: 
If the universe, either its content or in its entirety, was once (and still is) 
quantum, how can there be (apparently) classical structures now? 
\end{quotation}
We call this the ``cosmic measurement problem"
since it addresses the standard quantum measurement problem in the cosmological context.
 While many aspects of the cosmic measurement problem have been addressed in the 
literature, the perspective which we 
would like to add is that the problem is closely related to providing a physical 
basis for the (classical) FLRW model with a (classical) cosmic time 
 parameter.\footnote{Note the temporal aspect of the cosmic measurement
problem: Not only are classical structures less fundamental since they are 
derivable from quantum structures, but they are also {\em temporally} secondary to the
 original quantum state of the universe. Depending on the cosmic epoch of interest, 
various levels of the cosmic measurement
problem can be distinguished, for instance (see e.g. Kiefer and Joos 1999): (1) 
How to get a classical spacetime out of quantum spacetime? (2) How to get 
classical structures from quantum fields (in a classical spacetime background) --- 
for instance in an early inflationary universe? (3) How to get a measurement 
apparatus to show definite results (the standard measurement problem)?} 

Our point is that if cosmic time in the FLRW model is crucially 
dependent on a (prior) classical or classicalized behaviour of the 
material constituents of the universe, then one can hardly (assume a quantum 
fundamentalist view and) approach the cosmic measurement problem by 
asserting a gradual emergence of classicality framed {\em in terms of} 
a cosmic time.

An often attempted response to the cosmic measurement problem
is to
proceed via the idea of  decoherence. Within such an approach one may
imagine that quantum particles in the early universe (like particles in a bubble chamber) 
will move along `tracks' (instead of being wave functions spread out in space) --- 
due to the interaction of the quantum constituents with the environment (that is, 
the environment of all the other particles are constantly  
`monitoring' the particle wave function in question).
However, there are reasons to question whether decoherence has sufficient explanatory 
power for the quantum  fundamentalist (e.g. whether decoherence is sufficient
to explain the building up of a Weyl substratum).  
First, as is widely known, 
decoherence cannot by itself solve the measurement problem and 
explain the emergence of the classical world (see e.g. Landsman 2006).
Furthermore, as already indicated, if decoherence is to provide the 
classical structures (in the cosmological context), it cannot --- 
as is usually assumed in environmental induced decoherence --- be a process 
in (cosmic) time, insofar as classical structures (non-crossing world lines) are 
needed from the start to define cosmic time. Finally, a general worry about decoherence
 has been expressed e.g. by Anastopoulos (2002): ``\dots a sufficiently classical behaviour 
for the environment seems to be necessary if it is to act as a decohering 
agent and we can ask what has brought the environment 
into such a state ad-infinitum".\footnote{A further problem is that while 
the split between system and environmental degrees of freedom may be 
natural in earth-based experimental arrangements, it appears less obvious in the context of the early universe. Thus, while Kiefer and Joos (1999) appear to assume that various subsets of constituents in the universe can successively classicalize one another via decoherence (starting with the gravitational degrees of freedom), Anastopoulos (2002) points out that the environment/system splitting seems to be arbitrary in the context of general
relativity. }

Due to these limitations of the decoherence idea in the present context, 
the quantum fundamentalist is (in our view) still faced with the 
question of whether a comoving Weyl substratum can be constructed from 
(non classicalized) quantum constituents (wave functions).
 Apart from (but related to) the mentioned concerns about decoherence in this context, 
one may ask what `moves' according to  the quantum description?
>From the point of view of a Born interpretation, the wave function in quantum theory 
is not a real wave but rather a
probabilistic object. The evolution of a wave function $
\psi (\vec{\BFACE{x}}, t) $ therefore appears insufficient to provide a 
physical basis for the fluid particles comprising the Weyl substratum
since in the quantum description (in the Born interpretation) no physical object moves 
from a definite place
${\cal A}$ to another place ${\cal B}$.
Only a mathematical entity
$\psi(\vec{\BFACE{x}}, t)$ -- the symbolic representation of the quantum
system -- `moves'.\footnote{In our assessment, also the local space and time
concepts require a physical foundation in
terms of the material constituents (cf. the time-clock relation in Rugh and Zinkernagel 2009). 
In the quantum context (quantum mechanics as well as quantum field theory) we are therefore 
faced with an
interesting circularity: The wavefunction 
$\psi = \psi (\vec{\BFACE{x}},t) $ is defined on classical spacetime 
$(\vec{\BFACE{x}},t)$  but spacetime has in turn to be constructed with reference to the 
material building blocks, that is, to the wave functions
$\psi$ themselves.}

It is not obvious that this problem will be more tractable if
instead of one particle we have quantum systems composed of many constituents (see also 
Landsman 2006, p. 492). The early universe is envisaged to be 
described by a collection of interacting quantum fields. In general, these 
(matter and radiation) fields will be in an entangled state in which it is far from 
clear that individual particle trajectories are discernible. Thus, even with many 
constituents it is still not clear that something actually moves from one place 
to another. As a consequence, there may not be a well-defined
notion of particle trajectories (let alone non-crossing particle trajectories) 
in which case no Weyl substratum can be identified. In that situation, 
 no cosmic time can be defined and it thus seems difficult to maintain the quantum 
fundamentalist view of an early quantum `epoch' of the universe.

As a mathematical study, the FLRW model may be extrapolated back
arbitrarily close to $t = 0$. But as a physical model
nobody believes it `before' the Planck time.
As we have argued, however, there are interesting problems with 
establishing a physical basis for the FLRW model with a cosmic time, even before
(in a backward extrapolation from now) we might reach an `epoch' in which theories
of quantum gravity may come into play.

\section*{Acknowledgements}
We would like to thank audiences at talks in Copenhagen, Dortmund, Granada, Heidelberg, Leeds, Oxford, and Utrecht for helpful comments. We also thank
George Ellis and Erhard Scholz for comments on the manuscript. We finally thank the Spanish Ministry of Science and Innovation (Project FFI2008-06418-C03-02) for financial support.

\section*{References}

Charis Anastopoulos, ``Frequently asked questions about 
decoherence", in: {\em International Journal of Theoretical Physics} 41, 2002, 
pp. 1573-1590.

\vspace{.2cm}
\noindent
Gordon Belot, ``Dust, Time, and Symmetry", in: {\em British Journal for 
the Philosophy of Science}  56, 2005, pp. 255-91.

\vspace{.2cm}
\noindent 
Silvio Bergia and Lucia Mazzoni, ``Genesis and evolution of Weyl's 
reflections on de Sitter's universe", in: Hubert Goenner {\em et al} (Eds.), 
{\em The expanding worlds of general relativity}. Basel: Birkh\" auser 1999, pp. 325-343.  

\vspace{.2cm}
\noindent
Herman Bondi, {\em Cosmology} (2nd edition). Cambridge: Cambridge University Press 1960.

\vspace{.2cm}
\noindent
Dennis Dieks, ``Becoming, Relativity and Locality", in Dennis Dieks (Ed.), 
{\em The Ontology of Spacetime}. Amsterdam: Elsevier 2006, pp. 157-176.

\vspace{.2cm}
\noindent
J\" urgen Ehlers, ``Contributions to the Relativistic Mechanics of Continuous
Media" (1961), translated in: {\em General Relativity and Gravitation} 25, 1993, pp. 1225-1266.

\vspace{.2cm}
\noindent
J\" urgen Ehlers, ``Editorial note to: H. Weyl, On the general relativity theory", in: {\em General Relativity and Gravitation} 41, 2009, pp. 1655-1660.

\vspace{.2cm}
\noindent
George F. R. Ellis, ``Innovation, resistance and change: the transition to 
the expanding universe", in: Bruno Bertotti {\em et al} (Eds.), {\em Modern 
cosmology in retrospect}. Cambridge: Cambridge University Press 1990, pp. 97-113.

\vspace{.2cm}
\noindent
George F. R. Ellis, ``Contributions of K. G\" odel to relativity and cosmology", 
in: Petr Hajek (Ed.), {\em G\" odel '96: Lecture notes in logic 6}. 
Berlin: Springer-Verlag 1996, pp. 325-343.

\vspace{.2cm}
\noindent
George F. R. Ellis, ``83 years of general relativity and cosmology: progress and problems", 
in: {\em Classical and Quantum Gravity} 16, 1999, pp. A37-A75.

\vspace{.2cm}
\noindent
Stephen W. Hawking and George F. R. Ellis, {\em The large scale structure of space-time}. 
Cambridge: Cambridge University Press 1973.

\vspace{.2cm}
\noindent
Alexander Friedman, ``On the Curvature of Space" (1922), reprinted in: 
{\em General Relativity and Gravitation} 31, 1999, pp. 1991-2000.

\vspace{.2cm}
\noindent
Hubert Goenner, ``Weyl's contributions to cosmology", in Erhard Scholtz (Ed.) 
{\em Hermann Weyl's Raum -- Zeit -- Materie and a general 
introduction to his scientific work}. Basel: Birkh{\" a}user 2001, pp. 105-137.

\vspace{.2cm}
\noindent
Kurt G\" odel,  ``A remark about the relationship between relativity 
theory and idealistic philosophy" in: Paul A. Schilpp (Ed.), {\em Albert Einstein: 
Philosopher-Scientist}. La Salle, Illinois: Open Court, 1949, pp. 555-562.

\vspace{.2cm}
\noindent
Erich Joos {\em et al} (Eds.) {\em Decoherence and the Appearance of a 
Classical World in Quantum Theory}. Berlin: Springer 2003. 

\vspace{.2cm}
\noindent
Claus Kiefer and Erich Joos, ``Decoherence: Concepts and examples", in: 
Philippe Blanchard and Arkadiusz Jadczyk (Eds.), {\em Quantum future}. 
Berlin: Springer 1999,  pp. 105-128.

\vspace{.2cm}
\noindent
Andrzej Krasinski, {\em Inhomogeneous Cosmological Models}. Cambridge: Cambridge 
University Press 1997.

\vspace{.2cm}
\noindent
Nicolaas P. Landsman, ``Between classical and quantum", in: Jeremy Butterfield and 
John Earman (Eds.), {\em Handbook of the Philosophy of Science, Vol. 2: Philosophy of 
Physics}. Amsterdam: North-Holland, 2006, pp. 417-554. 

\vspace{.2cm}
\noindent
David B. Malament,  ``Classical general relativity", in: Jeremy Butterfield and John 
Earman (Eds.), {\em Handbook of the Philosophy of Science, Vol. 2: Philosophy of Physics}. 
Amsterdam: North-Holland, 2006, pp. 417-554.

\vspace{.2cm}
\noindent
Charles W. Misner, Kip S. Thorne, John A. Wheeler, {\em
Gravitation}. New York:  W. H. Freeman 1973.

\vspace{.2cm}
\noindent
Jayant V. Narlikar, {\em An Introduction to Cosmology}. Third Edition. Cambridge: 
Cambridge University Press 2002.

\vspace{.2cm}
\noindent
John D. North, {\em The measure of the universe -- A history of
modern cosmology}. New York: Dover 1990. 

\vspace{.2cm}
\noindent
Phillip J. Peebles, {\em Principles of physical cosmology}. Princeton: 
Princeton University Press 1993.

\vspace{.2cm}
\noindent
Amal K. Raychaudhuri, {\em Theoretical Cosmology}. Oxford: Clarendon Press 1979.

\vspace{.2cm}
\noindent
Howard P. Robertson, ``Relativistic Cosmology", in: {\em Reviews of Modern Physics} 5, 
1933, pp. 62-90.

\vspace{.2cm}
\noindent
Svend E. Rugh and Henrik Zinkernagel, ``On the physical basis of cosmic time", in: 
{\em Studies in History and Philosophy of Modern Physics} 40, 2009, pp. 1-19. 

\vspace{.2cm}
\noindent
Robert M. Wald, {\em General Relativity}. Chicago: The University of Chicago Press 1984.

\vspace{.2cm}
\noindent
Steven Weinberg, {\em Gravitation and Cosmology}. New York: John Wiley \& Sons 1972.

\vspace{.2cm}
\noindent
Hermann Weyl, ``On the general relativity theory" (1923), reprinted in: {\em General 
Relativity and Gravitation} 41, 2009, pp. 1661-1666.

\vspace{.2cm}
\noindent
Hermann Weyl, ``Redshift and Relativistic Cosmology", in:
{\em The London, Edinburgh, and Dublin Philosophical Magazine
and Journal of Science}  9, 1930, pp. 936-943.

\end{document}